\documentclass[12pt,preprint]{aastex}

\slugcomment{accepted for publication in ApJ}

\shorttitle{Supermassive Black Hole Binary}
\shortauthors{Matsui et al.}

\begin{document}

\title{Effects of A Supermassive Black Hole Binary on A Nuclear Gas Disk}

\author{Hidenori Matsui \altaffilmark{}}
\affil{Division of Physics, Graduate School of Science,
Hokkaido University, Sapporo 060, Japan}
\email{hidenori@astro1.sci.hokudai.ac.jp}

\author{Asao Habe \altaffilmark{}}
\affil{Division of Physics, Graduate School of Science,
Hokkaido University, Sapporo 060, Japan}
\email{habe@astro1.sci.hokudai.ac.jp}

\and

\author{Takayuki R. Saitoh \altaffilmark{}}
\affil{National Astronomical Observatory of Japan,
Mitaka, Tokyo 181-8588, Japan}
\email{saitoh.takayuki@nao.ac.jp}

\begin{abstract}
We study influence of
a galactic central supermassive black hole (SMBH) binary
on gas dynamics and star formation activity
in a nuclear gas disk
by making three-dimensional Tree+SPH simulations.
Due to orbital motions of SMBHs,
there are various resonances
between gas motion and the SMBH binary motion.
We have shown that these resonances create
some characteristic structures of gas
in the nuclear gas disk,
for examples,
gas elongated or filament structures,
formation of gaseous spiral arms,
and small gas disks around SMBHs.
In these gaseous dense regions,
active star formations are induced.
As the result,
many star burst regions are formed in the nuclear region.
\end{abstract}


\section{Introduction}

In the recent high-resolution observations
(e.g., $Chandra$ $X-ray$ $Observatory$),
the evidence of supermassive black hole
(here after we use SMBH for them) binaries
are shown in several galaxies,
e.g., in NGC 6240
\citep{kom03}, Arp 220 \citep{cle02},
M83 \citep{sak03, mas05},
and 3C 66B \citep{sud03}.
Particularly,
NGC 6240 has been well observed in the wide range of wave length
\citep{tac99,kom03}.
In the high-resolution X-ray observation
by $Chandra$ $X-ray$ $Observatory$,
strong two peaks of hard X-ray are detected in the galactic center
and this is the strong evidence of a SMBH binary \citep{kom03}.
By the radio continuum, the near-infrared,
and the soft X-ray observations of this galaxy,
nuclear star burst has been indicated
\citep{lir02, pas04}.
Nuclear gas rich disks have been observed
around the SMBH binary in NGC 6240 \citep{tac99}
and Arp 220 \citep{sco97, sak99}.

The SMBH binaries in galactic central regions
are expected to be formed in merging process
of galaxies each of which has a SMBH
in its galactic center.
After galaxies merge,
SMBHs sink into the center of the merging galaxy
by dynamical friction between SMBHs and field stars
\citep{ebi01, esc04, esc05}.
These SMBHs will form a SMBH binary
and finally merge due to emission of gravitational wave
\citep{mat04, eno04, esc04, esc05}.
In the process,
\citet{esc04, esc05} have shown
an important role of dynamical interaction
between SMBHs and gas,
especially in very gas rich regions.

\citet{kaz05} have made simulations of the merging process
of two disk galaxies with SMBHs at each galactic center.
Their results indicate that
much gas flows into the center of merging galaxy
and star burst is triggered.
They also show that
in the process,
a SMBH binary is formed in the center of merging galaxy
and
a nuclear gas disk with radius of $1-2$ kpc
is formed at the galactic center.
In the gas disk,
effects of the SMBH binary on a nuclear gas disk
are not studied yet
and it may be very important for star formation
in the galactic center.

Since the gravitational potential of a SMBH binary has
non-axisymmetric component,
we expect that
gas motion in a nuclear gas disk is strongly influenced
by a SMBH binary.
Change of gravitational potential due to orbital motions of SMBHs
may induce resonance phenomena in the disk
as in barred galaxies.
In barred galaxies,
gas motion is influenced
by resonances \citep{ath92}.
It was indicated that
these resonances trigger nuclear star burst in the barred galaxies
\citep{fuk91,wad92,elm94,wad95,fuk98}.
In a SMBH binary case,
similar resonances may trigger active star formation
in the nuclear gas disk.
Moreover,
it is interesting
that the SMBH binary may yield some peculiar gaseous features
in the nuclear disk
which can be used as an evidence of a SMBH binary.

In this paper,
we study the influence
of a galactic central SMBH binary
on gas motion in a nuclear gas disk
by hydrodynamic simulations
by using 3-dimensional Tree+SPH code.
We show that the resonances due to a SMBH binary
trigger formation of gas concentrations in the disk
and as the results star formation rate increases.

In section 2,
we present our simulation model.
In section 3,
we present the results of our simulations.
In section 4,
we summarize our results and give some discussions.

\section{Simulation model}

To study the influence of a galactic central SMBH binary
on a nuclear gas disk,
we simulate motion of gas and SMBHs
in a model galaxy
by using Tree+SPH code.

\subsection{Model galaxy and a SMBH binary}

We assume that a model galaxy of which gravitational potential
in the nuclear region
is similar to NGC 6240.

We assume a spherical stellar mass distribution
which is consistent with the observed rotation curve
in the inner region of the galaxy.
The rotation curve of CO gas is obtained
in the galactic central region
by \citet{tac99}.
We adopt the King model for the stellar component
\begin{equation}
 \rho_{\rm star} (r) =
 \frac{\rho_{\rm star,0}}{\big( 1 + r/r_{\rm star,0} \big)^{3/2}},
\end{equation}
where $\rho_{\rm star,0} = 15 M_{\odot}$ pc$^{-3}$
and $r_{\rm star,0} = 0.5$ kpc,
which is consistent with the observation.

In our simulations,
we set a SMBH binary in the central region of this model galaxy.
Each SMBH is modeled by the Plummer potential
in order to avoid numerical singularity in its gravitational potential.
Our model parameters of the SMBH binary
are shown in Table~\ref{tbl-1}.
We assume two cases for the masses of SMBHs.
Since masses of nuclei in NGC 6240 are estimated
as about $5\times 10^8 M_{\odot}$ \citep{tac99},
in the first case each mass of SMBHs
is $5\times 10^8 M_{\odot}$
and in the second case each mass of SMBHs is
$1\times 10^8 M_{\odot}$.
We assume four cases
for the initial orbits of SMBHs.
One is the case in which both SMBHs move initially
in a circular orbit
and the others are the cases
in which both SMBHs move initially
in elliptical orbits.
The moving direction of SMBHs in each case is same with
rotation of the gas disk.
We assume that
initial distance of each SMBH is $350$ pc
from the galactic center
and initial positions of SMBHs are
at the opposite side of the galactic center.
The separation between SMBHs is about $750$ pc in NGC 6240
\citep{tac99}.

Initially the nuclear gas disk is assumed to be uniform density.
Its radius is $2$ kpc.
Similar size of a gas disk is obtained by galaxy merger simulations
with SMBHs \citep{kaz05}.
Its thickness is $500$ pc,
and its temperature is $10^3$ K.
The mass of gas within $2$ kpc is $2\times 10^9$ $M_{\odot}$.
The gas disk rotates with circular velocity
and is in gravitational equilibrium in the model galaxy.

\begin{deluxetable}{lcccc}
\tabletypesize{\scriptsize}
\tablecaption{Parameters of our simulations \label{tbl-1}}
\tablewidth{0pt}
\tablehead{
\colhead{Model} & \colhead{Mass of SMBH1 [$M_{\odot}$]}
& \colhead{Mass of SMBH2 [$M_{\odot}$]}
& \colhead{Eccentricity of SMBH's orbit}
& \colhead{Semi-major axis of SMBH orbit}
}
\startdata
0 & --- & --- & --- & --- \\
1 & $5\times 10^8$ & $5\times 10^8$
& 0.00 & $350$ pc
\\
2 & $5\times 10^8$ & $5\times 10^8$
& 0.67 & $350$ pc
\\
3 & $5\times 10^8$ & $5\times 10^8$
& 0.82 & $350$ pc
\\
4 & $5\times 10^8$ & $5\times 10^8$
& 0.94 & $350$ pc
\\
5 & $1\times 10^8$ & $1\times 10^8$
& 0.00 & $350$ pc
\\
6 & $1\times 10^8$ & $1\times 10^8$
& 0.64 & $350$ pc
\\
7 & $1\times 10^8$ & $1\times 10^8$
& 0.82 & $350$ pc
\\
8 & $1\times 10^8$ & $1\times 10^8$
& 0.93 & $350$ pc
\enddata
\end{deluxetable}

\subsection{Effect of a SMBH binary on gas dynamics}

The gravity of a SMBH binary has a non-axisymmetric component
as in a bar potential.
In barred galaxies,
it has been shown that
resonances between gas motion
and a rotating non-axisymmetric gravitational potential
are important for gas dynamics.
In a weak barred potential,
the Inner Lindblad, the Corotation, and Outer Lindblad resonances
are shown by the epicycle approximation.
These resonances play an important role in
formation of gaseous ridges and spiral arms
in barred galaxies.
(e.g., \citet{ath92}).

If SMBHs rotate in circular orbits,
gravitational potential will change with its pattern speed,
$\Omega _{BH}$,
which is orbital angular velocity of the SMBH.
Then, we expect the resonances
between gas motion and SMBH binary motion.

When orbits of SMBHs are elliptical,
the major axes of the elliptical orbits shift with time.
In this case,
due to the shift of major axis of each elliptical orbit,
time variation of gravitational potential is different
from the circular orbit case.
The angular velocity of this shift,
$\Omega_{P}$,
may be slower than $\Omega _{BH}$.
In this case,
we expect another resonances similar to
the Lindblad and Corotation resonances.
However,
time variation of this gravitational potential is very complicated.
Therefore,
we need to make numerical simulations
of gas including a SMBH binary.

\begin{figure}
\plotone{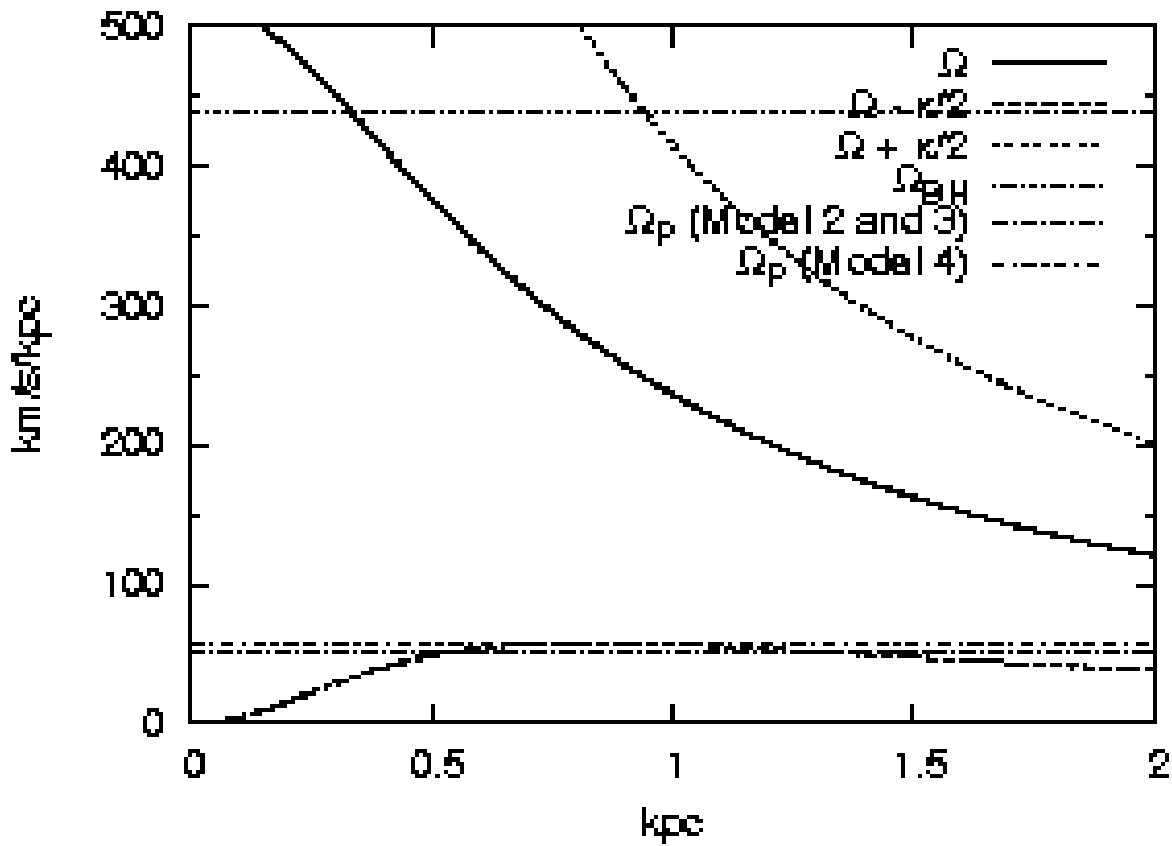}
\caption{The frequencies as a functions of radial distance
in the model galaxy.
We denote $\Omega _{BH}$ as the angular speeds of SMBHs
in the circular orbit of radius $350$ pc,
and $\Omega_p$ as the angular speed of the orbital precessions.
In our model,
$\Omega _{BH} = 438$ km s$^{-1}$ kpc$^{-1}$,
and $\Omega _{P} = 52, 52,$ and $57$ km s$^{-1}$ kpc$^{-1}$
for Model 2, 3, and 4,
respectively.\label{res}}
\end{figure}

We show the resonances in our simulation models in Fig.~\ref{res}.
Fig.~\ref{res} shows
the angular frequency of circular rotating gas motion
in the model galaxy potential,
$\Omega _{BH}$,
and $\Omega _{P}$.
In the Model 1,
SMBHs move in a circular orbit.
In this model,
we expect the Corotation and Outer Lindblad resonances
between gas motion and the SMBH binary motion
for $\Omega _{BH}$.
In the Model 2, 3, and 4,
SMBHs move in the elliptical orbits.
In these models,
$\Omega _{P}$ is smaller than $\Omega _{BH}$ as shown in Fig.~\ref{res}.
We can expect resonances due to low $\Omega _{P}$.

\subsection{Numerical method}

We use Tree+SPH code with GRAPE-5
to simulate the motion of gas and the SMBH binary.
In the code,
we solve gravitational force of gas and SMBHs
by using the combination of Tree method \citep{app85}
and GRAPE-5 \citep{sug90},
and hydrodynamic evolution is solved
by the Smoothed Particle Hydrodynamics (SPH) method
\citep{luc77, gin77}.
The neighbor search is accelerated
by the combination of GRAPE and reordering method
\citep{sai03}.
We consider self-gravity of gas in the fixed stellar potential.
Radiative cooling, star formation,
and thermal heating from supernovae are also considered.
We assume the Salpeter initial mass function
for newly formed stars.
The motions of newly formed star particles are followed
in the fixed stellar gravitational force
and gravity of gas, SMBHs, and other newly formed stars.

A SPH kernel is defined by
\begin{equation}
W(x,h)= \frac{1}{4\pi h^3}
\left\{
\begin{array}{lr}
4-6x^{2}+3x^{3} & (0\leq x \leq1),  \\
(2-x)^3 & (1 \leq x \leq 2), \\
0 & (2 \leq x),
\end{array}\right.
\end{equation}
where $h$ is particle's smoothing length,
$x=r_{ij}/h_i$,
and
$r_{ij} = | \mbox{\boldmath $r$}_i - \mbox{\boldmath $r$}_j |$.
The equations of motion and energy of the $i$-th SPH particle are
\begin{equation}
 \frac{d\mbox{\boldmath $r$}_i}{dt}
= \mbox{\boldmath $v$}_i
,
\end{equation}
\begin{eqnarray}
\frac{d\mbox{\boldmath $v$}_i}{dt}
= - \sum_j m_j\Big(\frac{P_i}{\rho _i^2}
+\frac{P_j}{\rho _j^2} +\Pi_{ij}\Big)
\nabla W(x,h) \nonumber \\
-\nabla (\Phi _{\rm star}
+\Phi _{\rm SMBH} + \Phi _{\rm gas} )
,
\end{eqnarray}
\begin{eqnarray}
 \frac{du_i}{dt}
= \sum_j \Big(\frac{P_i}{\rho _i^2}
+\frac{1}{2}\Pi_{ij}\Big)
\mbox{\boldmath $v$}_{ij}
\cdot
\nabla W(x,h) \nonumber \\
+ \frac{\mathcal{H}_i - \Lambda _i}{\rho_i} ,
\end{eqnarray}
where $P_i$ and $\rho_i$ are the pressure and density,
$\Pi _{ij}$ is the artificial viscosity
of which viscous parameters are $\alpha =1$ and $\beta = 2$
\citep{mon83},
$\mathcal{H}_i$ is supernovae heating rate by newly born stars,
and $\Lambda_i$ is the radiative cooling function
of H/He for $T>10^4$ K
and molecular gas for $10$ K $<T<$ $10^4$ K \citep{spa97}.
The metal line cooling is not considered in our simulations.
We find that it does not affect our results,
since temperature of gas in our simulations is lower than $10^5$ K.
We employ the shear reduced technique in the artificial viscosity
\citep{bal95}.

Star formation algorithm is similar to one by \citet{kat92}
\citep{sai04},
but for higher density criterion.
If a SPH particle satisfies all following conditions,
(1) higher number density than that of typical CO cloud density
($n_H > 200$ cm$^{-3}$)
in order to avoid the star formation in the low gas density regions,
(2) the Jeans criterion,
and (3) collapsing regions ($\nabla \cdot \mbox{\boldmath $v$}< 0$),
then a SPH particle is changed to a collisionless star particle.
Star formation efficiency is $0.033$.

We use 50000 SPH particles.
Since we assume that
the total mass of the nuclear gas disk within $2$ kpc is
$2\times 10^9 M_{\odot}$,
the mass of each SPH particle is $4\times 10^4 M_{\odot}$.
The gravitational softening lengths of SPH particles and SMBHs
are $170$ pc and $50$ pc,
respectively.
We employ the second order leapfrog method for time integration.
We also calculate by using 100000 SPH particles
to confirm our results.

\section{Results}

We show the numerical results
of gas motion in a nuclear gas disk with a SMBH binary.
In the model galaxy without the SMBH binary,
gas evolution in the gas disk is rather quiet.
On the other hand,
in the model galaxy with the SMBH binary of $5\times 10^8 M_{\odot}$,
resonances due to the SMBH binary
induce large peculiar gas motion in the gas disk.
Especially,
in high eccentric orbit cases of the SMBH binary,
gas motion is influenced very much.
However,
in the model galaxy with the SMBH binary of $1\times 10^8 M_{\odot}$,
the effects of the SMBH binary are not strong.
This can be understood by the fact
that mass of the smaller SMBH is only $2.4$ \%
of dynamical mass of the model galaxy
within $500$ pc,
$4.1 \times 10^9 M_{\odot}$.
Therefore,
we show the results below in the case
of SMBHs of $5\times 10^8 M_{\odot}$.

\begin{figure}
\plotone{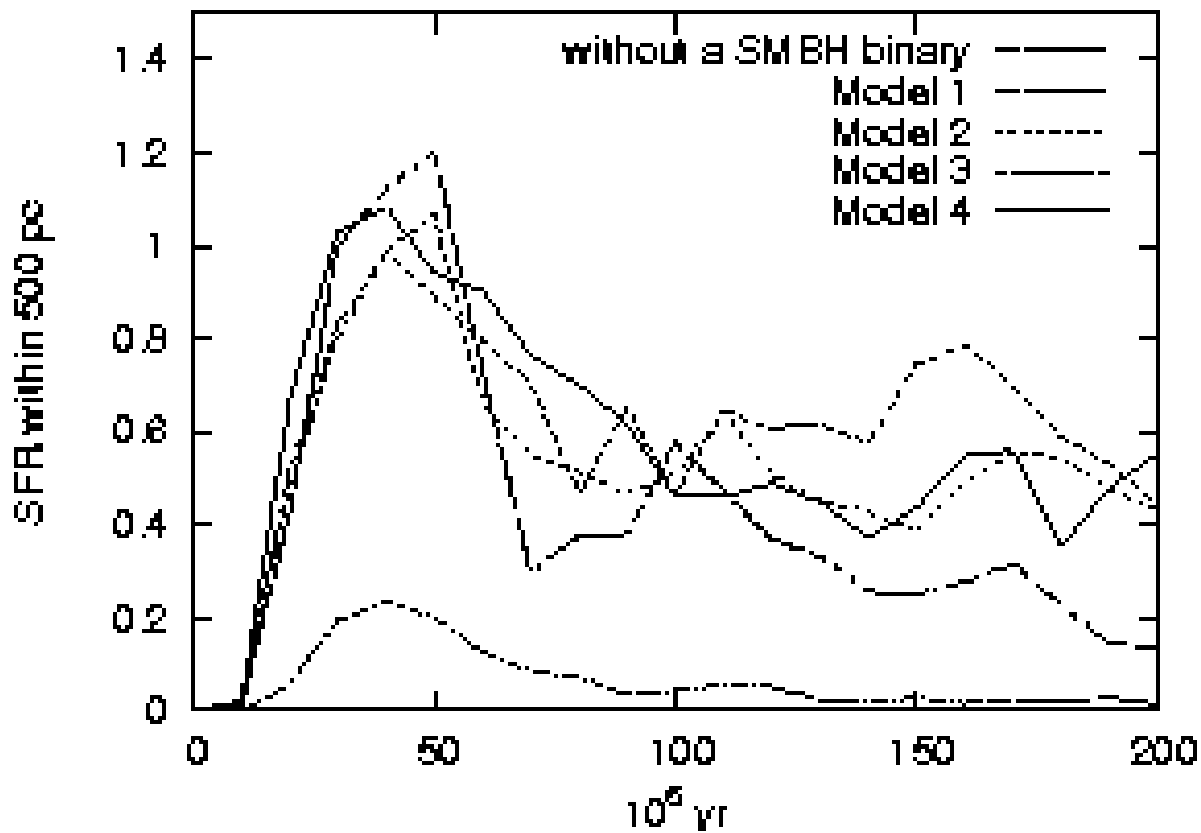}
\caption{The time variation of star formation rate within $500$ pc
($M_{\odot}$yr$^{-1}$)
for our Models.
The dot-dashed, dotted, short-dashed, long-dashed, and solid lines
show star formation rate in Model 0, 1, 2, 3, and 4,
respectively.
\label{sfr}}
\end{figure}

\subsection{The circular orbit case}

We show the result of the model
with the SMBH binary
in which each SMBH moves in its circular orbit
(Model 1).
In this case,
we expect that there are
the Corotation Resonance ($r_{\rm CR}\sim350$ pc)
and the Outer Lindblad Resonance ($r_{\rm OLR}\sim900$ pc)
as shown in Fig.~\ref{res} (see \S 2.2).

\begin{figure}
\plottwo{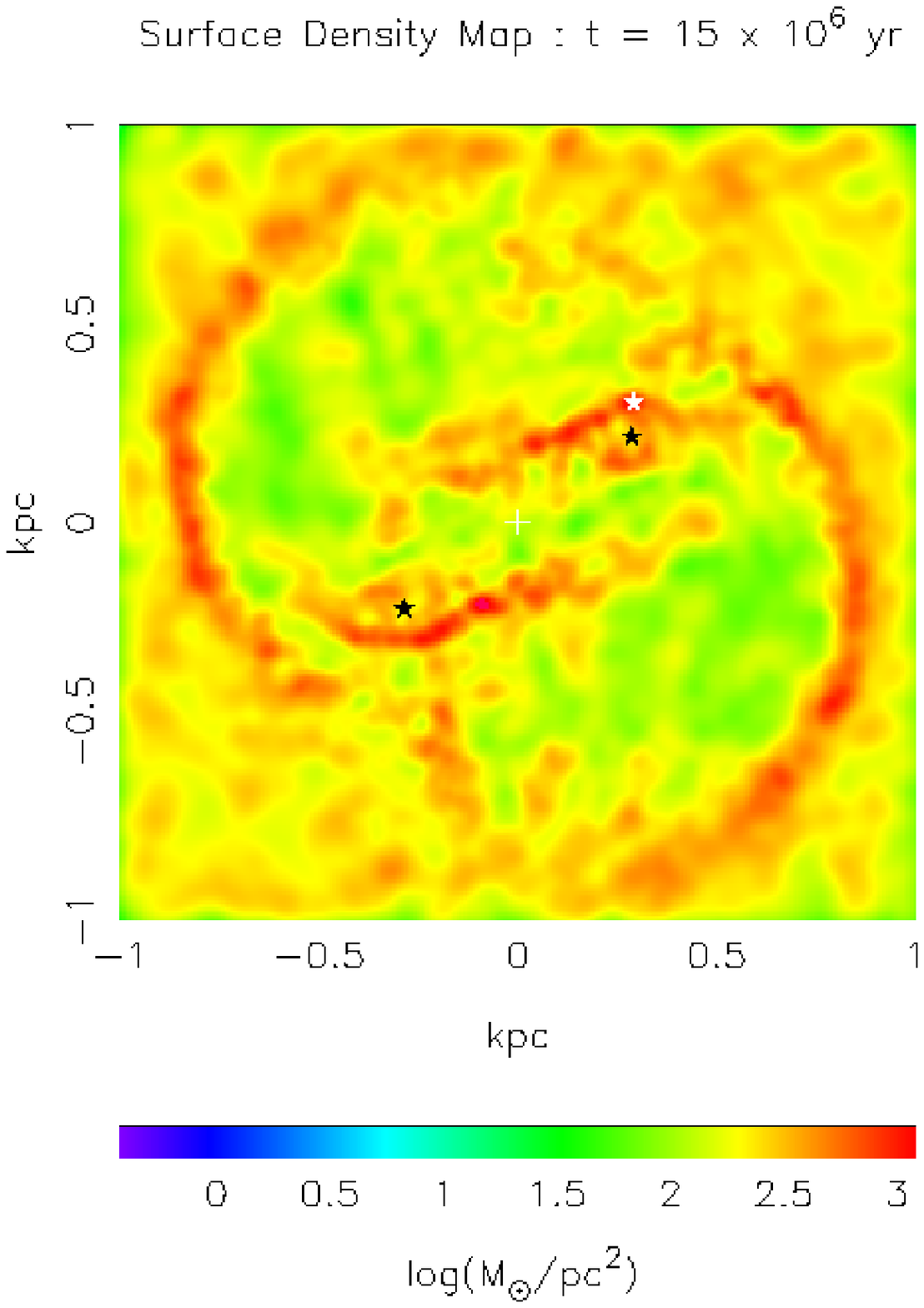}{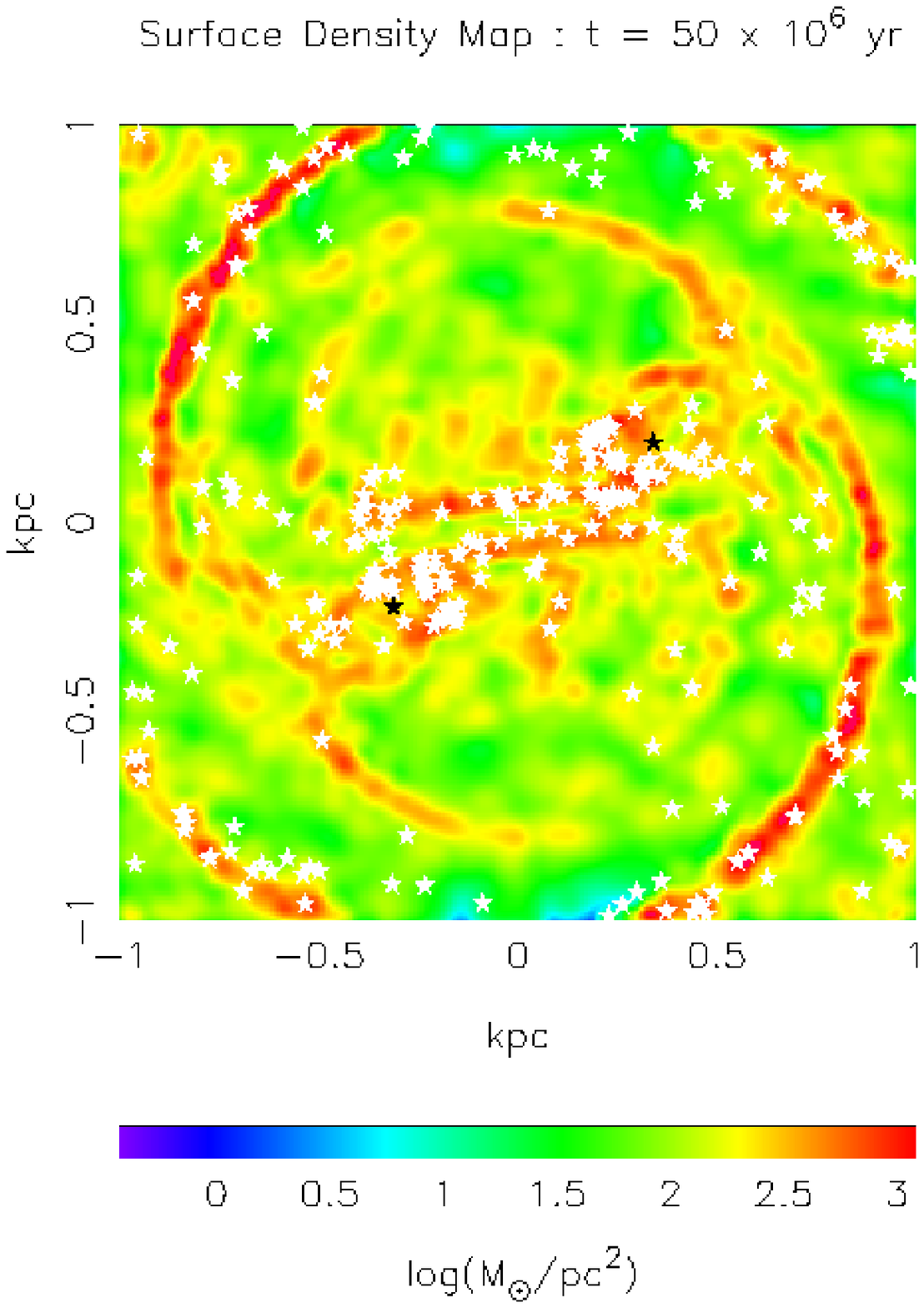}
\caption{The results of Model 1.
The left panel shows the result at $t=1.5\times 10^7$ yr
and the right panel shows the result at $t=5\times 10^7$ yr.
The positions of SMBHs are represented
by black stars.
Newly formed stars of which age is
from $5.4\times 10^6$ yr to $1 \times 10^7$ yr
are shown by symbols of a white star.
\label{result1}}
\end{figure}

Our numerical result shows that
SMBHs influence on gas motion of the nuclear disk
and excite active star formation in the nuclear region.
In the left panel of Fig.~\ref{result1},
we show the gas surface density and star formation sites
at $1.5\times 10^7$ yr in Model 1.
At that time,
the effects of resonances appear in gas distribution.
Gaseous ridge structures appear
at the upstream side of the SMBH binary.
Gaseous spiral arms are formed
in $700-900$ pc
where is the vicinity of radius of the Outer Lindblad Resonance.
This may be due to the Outer Lindblad Resonance.
These ridge and spiral arm structures resemble
those of barred galaxies
\citep{ath92}.
After the formation of ridges,
gas is accumulated into the ridges
and gas mass in the ridges increases with time.
The gaseous ridges change their shape
and an elongated gas rich region is formed
between the SMBHs
as shown in the right panel of Fig.~\ref{result1}.
The major axis of the elongated region is parallel to
the SMBH position angle
and its size of semi-major axis is about $500$ pc
and the minor axis is about $300$ pc.
In the elongated region,
gas becomes cool due to radiative cooling
and many dense clumps are formed
due to gravitational instability.
In these clumps,
star formation occurs,
since these clumps satisfy the star formation criteria
given in \S2.3.
As the result,
in these dense gas regions,
star formation becomes very active
as shown in the right panel of Fig.~\ref{result1}.
After the active star formation stage,
mass of gas decreases with time in the nuclear disk,
since much gas was spent by star formation.
These features are very different from the case
without a SMBH binary
in which gas surface density is axisymmetric,
gas concentration does not occur in the galactic center,
spiral arms are not formed,
and star formation is not active in the central region.

The time variation of star formation rate (SFR)
within $500$ pc radius is shown in Fig.~\ref{sfr}.
The active star formation has continued
from $2\times 10^7$ yr to $6\times 10^7$ yr
and the mean SFR in this period
is about $1 M_{\odot}$ yr$^{-1}$
which is much higher than that of the Model 0.
Duration time of the star bursts is less than $10^8$ yr.
After that,
SFR declines gradually.
This is because most gas transforms into stars
and then gas is deficient in the galactic central region.
After active star formation stage,
the total mass of newly formed stars is about $4.8\times 10^7 M_{\odot}$
within $500$ pc.
The mass is 34\% of initial gas mass within $500$ pc
which is about $1.4 \times 10^8 M_{\odot}$.

\subsection{The elliptical orbit cases}

We show the results of the models with a SMBH binary
in which each SMBH moves in the elliptical orbit
(Model 2, 3, and 4).
From Model 2 to 4,
we increase eccentricity of the SMBH orbits
as shown in Table~\ref{tbl-1}.
In these cases,
the major axis of the elliptical orbit shifts with time.
As shown in \S 2.2,
the shift may excite additional resonance for $\Omega _{P}$
which is angular velocity of the shift.
We show that
in these cases
the gas morphology and the star formation sites
are different
from the circular orbit case (\S 3.1).

\begin{figure}
\plottwo{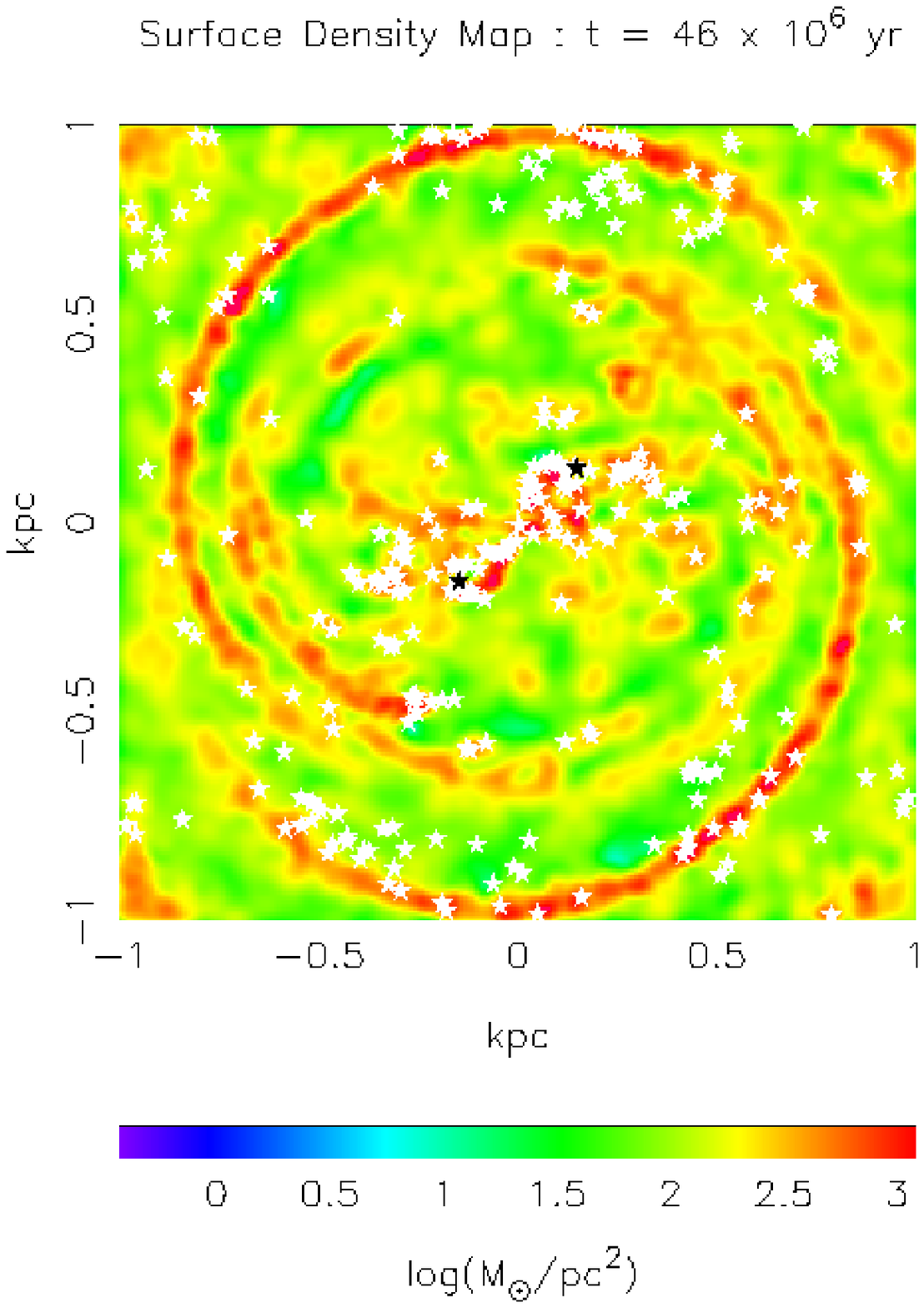}{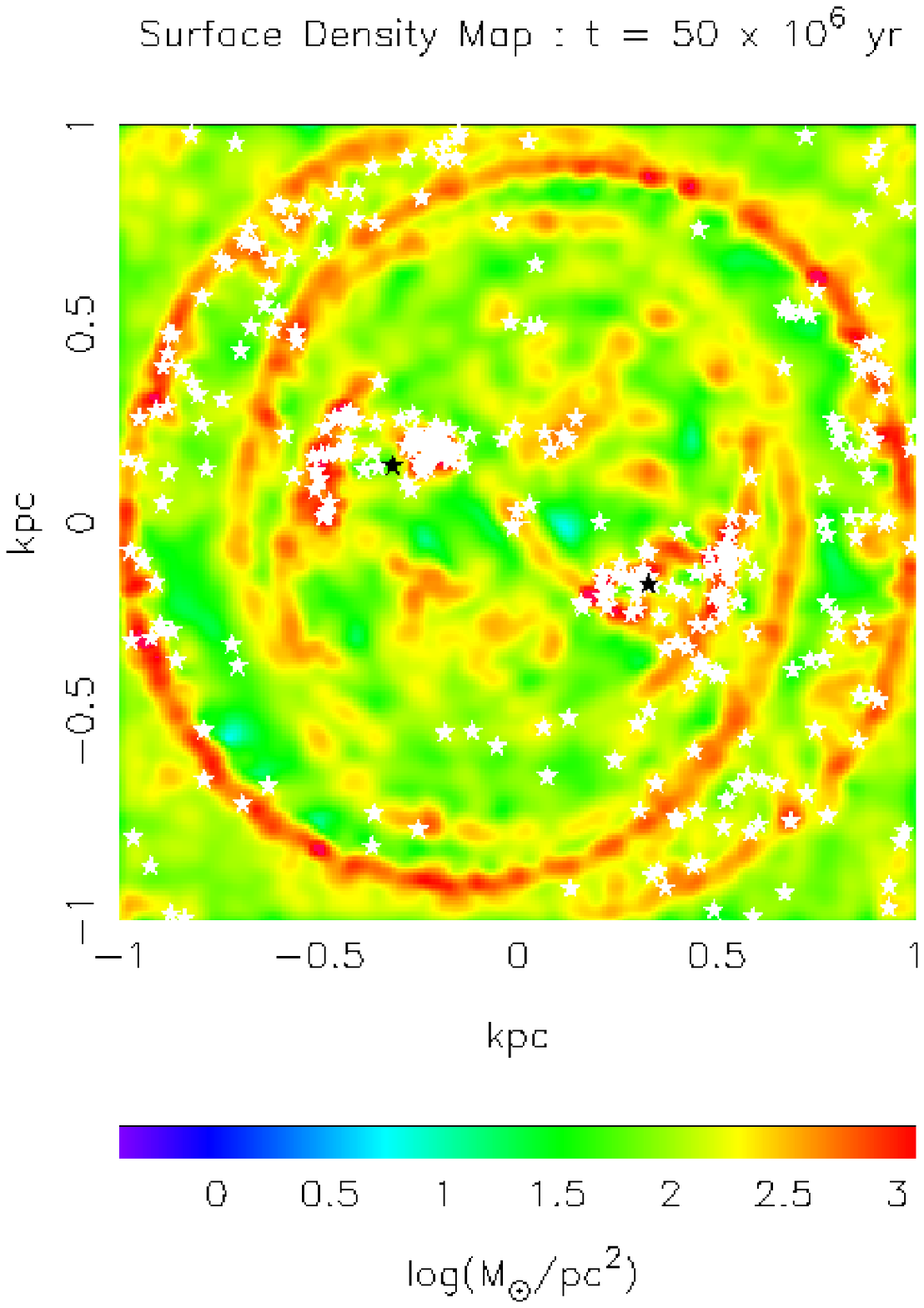}
\caption{
The same as Fig.~\ref{result1} but for Model 3.
\label{result3-1}}
\end{figure}

We show the result of Model 3
in which eccentricity of the SMBH orbit is about $0.82$.
At first,
the gas ridge structures are formed
and are parallel to the major axis of the SMBH binary.
Spiral arms are also formed
around the binary as in Model 1.
When the SMBHs are close each other,
a bar-like dense gas structure is formed
between SMBHs due to increase of
the non-axisymmetric component
of the gravitational potential of the SMBH binary
as shown in the left panel of Fig.~\ref{result3-1}.
After that,
the dense gas bar is elongated by the gravity of receding SMBHs
and a gas filament structure is formed between SMBHs.
In this process,
a part of gas in the gas filament is captured by the gravity of SMBHs.
As the results,
gas becomes dense
in the filament, spiral arms, and dense regions around SMBHs.
In these dense regions,
gas becomes cool due to radiative cooling
and star formation becomes active
as shown in the right panel of Fig.~\ref{result3-1}.
The regions of active star formation are more compact than Model 1,
although the SFR within $500$ pc is similar to Model 1
as shown in Fig.~\ref{sfr}.

\begin{figure}
\plotone{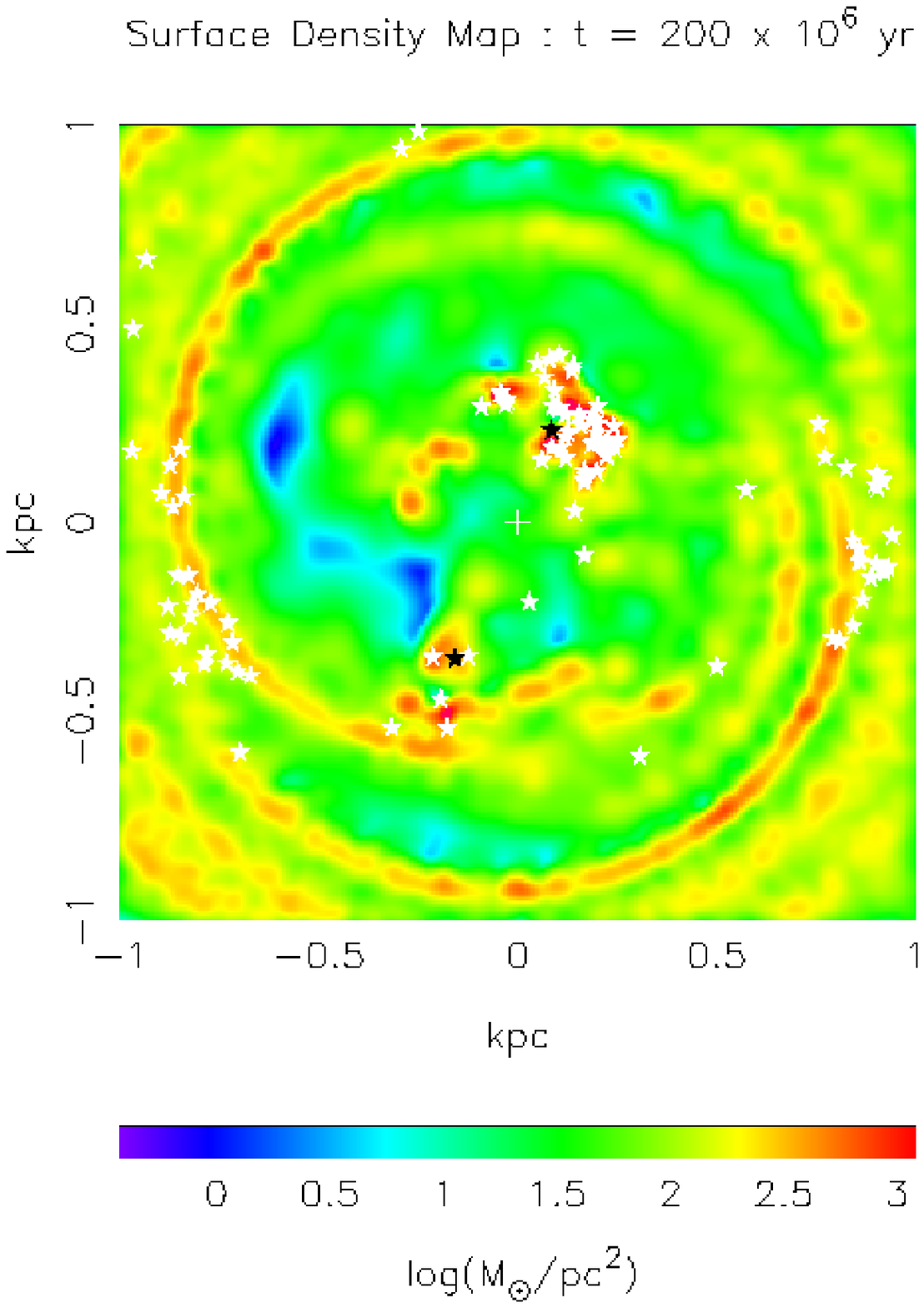}
\caption{
The same as Fig.~\ref{result1} but for Model 3.
\label{result3-2}}
\end{figure}

After several orbital rotations of SMBHs,
small gas disks are formed around SMBHs
and active star formation occurs in the disks
as shown in the Fig.~\ref{result3-2}.
The total of the gas and stellar mass
in the small disk around each SMBH is
about $3\times 10^7 M_{\odot}$
at $2\times 10^8$ yr.
Such gas disks are not formed around SMBHs in the Model 1.
The process of the gas disk formations around SMBHs is as following.
After SMBHs pass through each other,
the dense gas component is captured by the SMBHs
and dense gas distributes around SMBHs
as shown in the right panel of Fig.~\ref{result3-1}.
A part of accumulated gas around SMBHs evolves to
the small gaseous disks around SMBHs.
In the circular orbit case of SMBHs,
SMBHs can't capture much gas,
probably because SMBHs are not close each other
and the dense gas filament is not formed.
Therefore,
much gas can't be accumulated into the regions around SMBHs
by the gravity of SMBHs
and gas disks are hardly formed around SMBHs
in Model 1.
The massive gas disks
and the active star formation around SMBHs
are the typical features of elliptical orbit case of SMBHs.

\begin{figure}
\plottwo{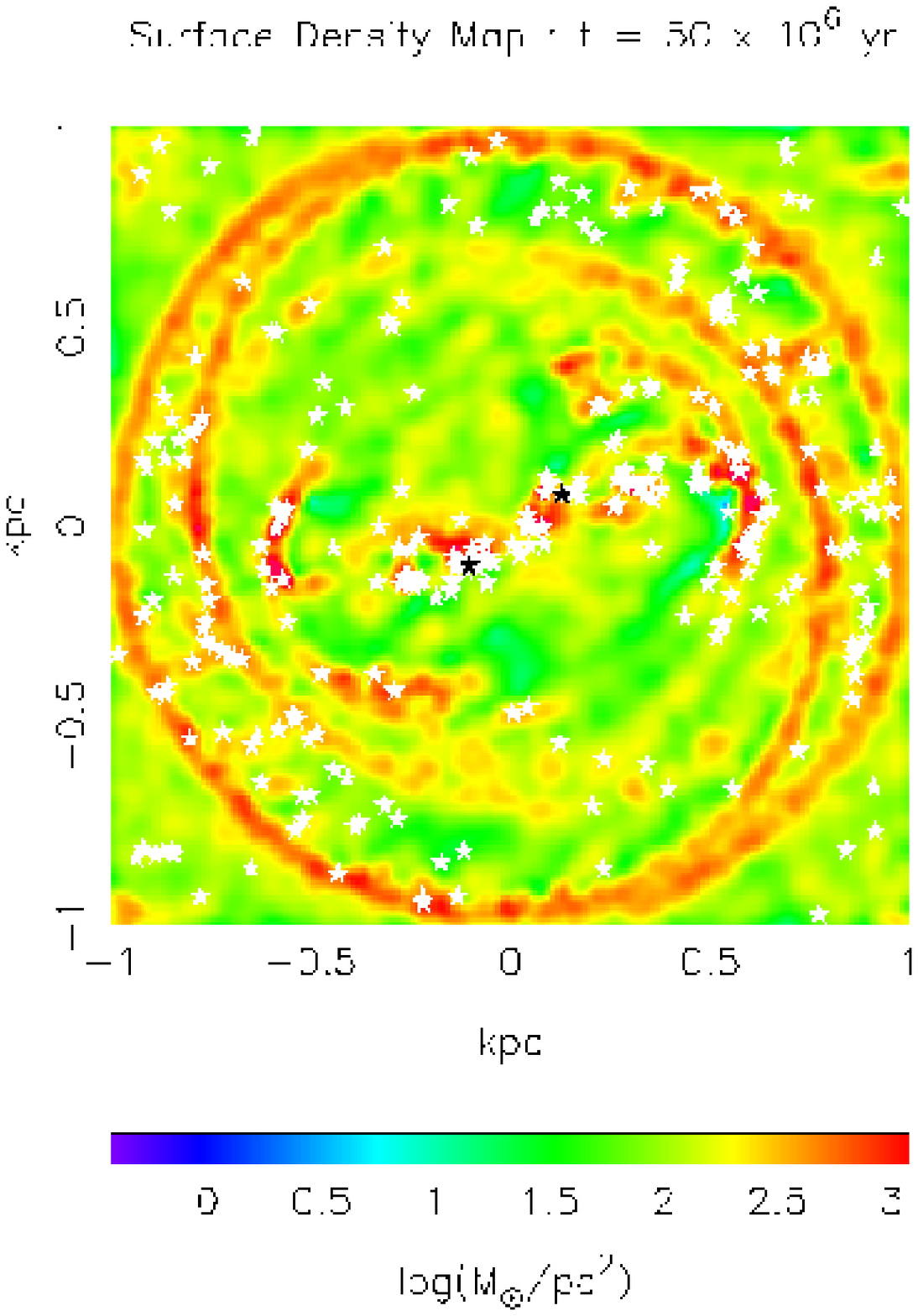}{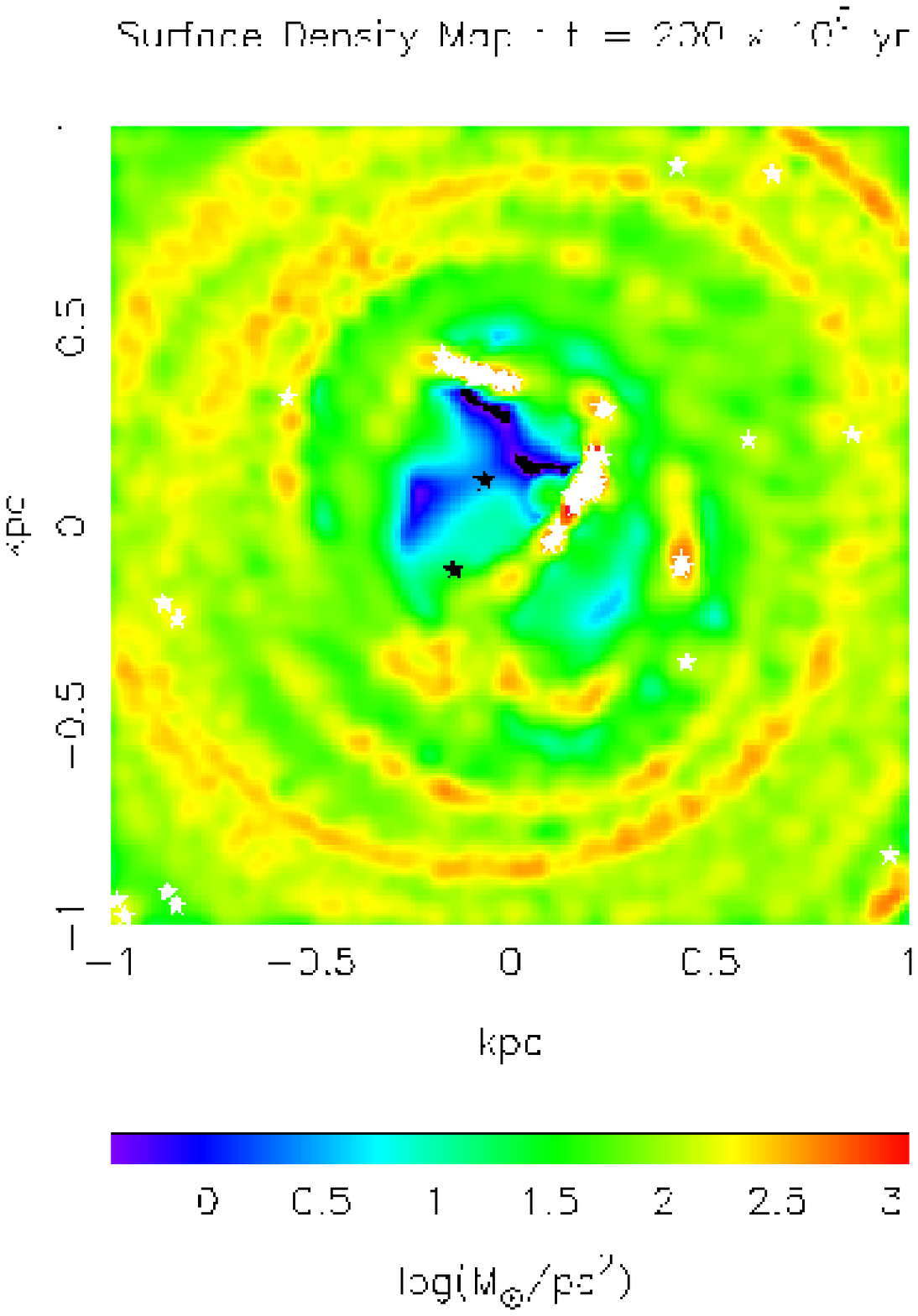}
\caption{
The same as Fig.~\ref{result1} but for Model 4.
\label{result4}}
\end{figure}

Next,
we show the result of Model 4
in which eccentricity of both SMBHs' orbits
is very high and about $0.93$.
The left panel of Fig.~\ref{result4} shows
the gas surface density and star formation sites
in the model
at $t=5\times 10^7$ yr
during active star formation stage.
The figure shows that
the evolution of gas and star formation sites
are similar to Model 3.
Fig.~\ref{sfr} shows that
SFR is also similar to Model 3.

After several orbital rotations of SMBHs,
gas disks are also formed around SMBHs
as in Model 2 and 3
as shown in the left panel of Fig.~\ref{result4}.
However,
after about $1.5\times 10^8$ yr,
the gas disks around SMBHs are destroyed
as shown in the right panel of Fig.~\ref{result4}.
This is induced
by the strong tidal force due to another SMBH,
when SMBHs are close each other.

The tidal force also affects
the orbital angular momentum of each SMBH
and it may be important for coalescence of SMBHs.
When the tidal force destroys the disks around SMBHs,
the destroyed part of gas disks gains the angular momentums from SMBHs.
As the result,
each SMBH loses its orbital angular momentum.
In the other models
the angular momentums are not decreased as much as Model 4.
This may be due to the fact
that the destruction of disks around SMBHs
by the tidal force
don't occur in these models.

\begin{figure}
\plottwo{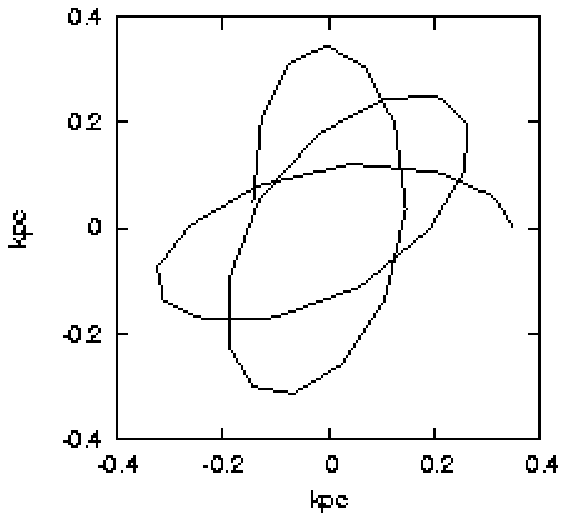}{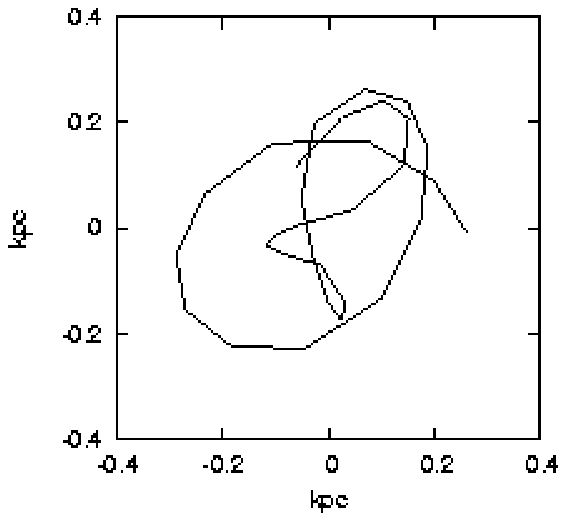}
\caption{
The orbit of the SMBH in Model 4.
The left panel shows the orbit from $t=0$ to $3\times 10^7$ yr
and the right panel shows the orbit
from $t=1.7\times 10^8$ to $2\times 10^8$ yr.
\label{orbit}}
\end{figure}

Since large part of orbital angular momentum of SMBHs
decreases with time,
the violent irregular motions of SMBHs appear
in Model 4.
In the early stage,
SMBHs move in their original elliptical orbits
similar to the other models
as shown in the left panel of Fig.~\ref{orbit}.
After about $3\times 10^7$ yr,
since SMBHs lose its angular momentum
by the tidal interaction,
the motions of SMBHs begin to deviate
from their original elliptical orbits.
Due to the deviation,
position of each SMBH becomes asymmetry to the galactic center.
Thus,
after about $1.0\times 10^8$ yr,
each angular momentum of SMBHs is exchanged with each other
by the gravitational force of another SMBH,
since the force is not directed to the galactic center
and is dominant near the galactic center.
As the result,
after about $1.5\times 10^8$ yr,
large irregular motions of SMBHs are induced
as shown in the right panel of Fig.~\ref{orbit}.
We check this result
by a simulation with half time step for the motions of SMBHs.
Similar violent irregular motions of SMBHs are obtained
in this test simulation.
We also check this result
by a simulation of a SMBH binary without gas.
In the test simulation,
SMBHs continue to move in their original elliptical orbits.
We can conclude that
the irregular motion in the elliptical orbit case is not artifact.

\section{Summary and discussion}

We study influence of the galactic central SMBH binary
on gas dynamics in the nuclear gas disk
by numerical simulations.
We calculate various cases for initial SMBH orbits
which are the circular orbit case and the elliptical orbit cases.
We have shown that
in the all cases,
the SMBH binary has large
influence on gas motion in the gas disk.
The SMBH binary induces some resonances on gas motion
in a nuclear gas disk.
Due to these resonances,
various dense gas structures are formed in the nuclear gas disk
and gaseous spiral arms are formed
near the vicinity of Outer Lindblad Resonance.
In these dense gas regions,
star formation becomes very active.
In the high eccentric orbit cases of the SMBH binary,
gaseous narrow filaments and dense clump structures are developed well,
and active star formation occurs in these regions.
These features can be
strong evidence of existence of a SMBH binary.
It is very interesting to compare these features
with very high-resolution observations of galaxies
which are proposed to have a SMBH binary.
Dense gas structures and distribution of star formation cites
will inform us dynamical state of the SMBH binary.
It should be noticed that
these features are not appeared in the case
of which the SMBH mass is smaller than about $1\times 10^8$ $M_{\odot}$
in our model.
We note that
when we compare whole SFR of ultra luminous infrared galaxies,
the SFRs induced by the SMBH binary are not high.
However,
the SFR is as high as nuclear star bursts in nearby star burst galaxies.

In our simulations,
gaseous ridges are formed by shocks.
In the elliptical cases of SMBHs,
there are the collisions between gas clumps in the galactic center.
Shocks are exited by these collisions of gas clumps.
H$_2$ emission line is expected to be excited in the shocks.
\citet{van93} and \citet{sug97}
have observed bright H$_2$ emission line
in the galactic central region of NGC 6240
and they conclude that the H$_2$ emission is excited by shock.

From our numerical simulations,
small gas disks are formed around SMBHs
in the elliptical orbit cases.
Such gas disks around SMBHs
have been observed in Arp 220
which has the double nuclei in the galactic center
\citep{sak99}.
In our simulations,
star formation is very active in the small gas disks around SMBHs.
The active star formation in the gas disks around SMBHs
may correspond to radio continuum sources observed
around SMBHs in NGC6240 \citep{tac99}.
$Chandra$ $X-ray$ $Observatory$ observed hard X-ray
from double nuclei in NGC 6240.
It is possible to excite AGN activity by
gas accretion onto SMBHs in the small gas disks.
If AGN activities are highly excited
and AGN feedback becomes very strong,
active star formation will be quenched \citep{mat05}.

Since active star formation occurs in very compact regions
in the highly eccentric elliptical orbit cases,
we expect that these newly formed stars concentrate
in compact massive star clusters.
If these star clusters interact gravitationally
with the SMBH binary,
the interaction may induce losing
of orbital angular momentum of SMBHs
due to the unstableness of three body problem
in which SMBHs and the star cluster interact with each other
and ejection of the star cluster occurs.
If these star clusters are massive enough,
this process may be very effective
and the binary can evolve to a more tightly binding state.
This process may have an important role in merging process of SMBHs.

\citet{esc05} studied
the effect of hydrodynamic drag force by dense gas
on evolution of a SMBH binary
by numerical simulations.
They have shown that
after SMBHs gradually fall into the galactic central dense gas region
by the dynamical friction,
effect of hydrodynamical drag becomes very effective
in the central region.
They suggested that finally SMBHs can be close enough to merge
by the hydrodynamic effect.
In their simulations,
they don't consider
effects of radiative cooling and star formations on gas.
By the effect of radiative cooling,
many dense clump structures will be formed
and those distribution is more complicated.
The dense clumps may interact with SMBHs
and play an important role in the coalesce of SMBHs.
Moreover,
active star formation will occur in the clumps
and gas mass will decrease in the galactic central region.
After the active star formation,
it is not clear
that in the galactic center
gas remains enough for hydrodynamic interaction with SMBHs or not.
It is needed to make simulations of evolution of a SMBH binary
in more realistic model.

In our simulations,
we didn't consider the dynamical friction
between field stars and SMBHs.
The dynamical friction induces decay of orbital radius of SMBHs.
If timescale of dynamical friction is larger
than timescale of rotation motion of SMBHs,
the resonances between SMBH motions and gas motion will be effective.
In this case,
similar process appeared in our simulations will occur.
On the other hand,
if the dynamical friction timescale is smaller
than the rotation timescale,
orbits of SMBHs shrink very rapidly
and the resonance phenomena are not important.

We have assumed that
initially gas disk is a circularly rotating.
However, since a galaxy with a SMBH binary
is expected to be formed due to merging of galaxies with SMBHs,
gas motion is more complex in merging galaxies.
To simulate more realistic evolution of a galaxy with a SMBH binary,
we will study merging process of galaxies with SMBHs.
In this process,
radiation drag \citep{kaw05}
and the influence of AGN feedback \citep{spr05}
should be considered.

\acknowledgments
We thank Professor Masayuki Fujimoto, Professor Kazuo Sorai,
Dr. Tamon Suwa, Dr. Kimitake Hayasaki, and Mr. Junya Itou
for helpful discussions.
This work has been supported by
Grant-in-Aid for the 21st Century COE Scientific Research Programme
on ``Topological Science and Technology''
from the Ministry of Education, Culture, Sport, Science,
and Technology of Japan (MECSST),
in part by Grant-in-Aid for Scientific Research
(14340058)
of Japan Society for the Promotion of Science,
and in part by Hokkaido University Grant Program
for New Fusion of Extensive Research Fields.

\clearpage

\clearpage

\end{document}